# The monoenergetic approximation in stellarator neoclassical calculations

Matt Landreman

Massachusetts Institute of Technology, Plasma Science and Fusion Center,
167 Albany St., Cambridge, MA 02139, USA
E-mail: landrema@mit.edu

**Abstract**

In the standard "monoenergetic" approach to numerical calculation of stellarator neoclassical transport, to expedite computation, ad-hoc changes are made to the kinetic equation so speed enters only as a parameter. Here we examine the validity of this approach by considering the effective particle trajectories in a model magnetic field. We find monoenergetic codes systematically under-predict the true trapped particle fraction, with the error in the trapped ion fraction being of order unity when the electric field is large, suggesting some results of these codes may be unreliable in this regime. This inaccuracy is independent of any errors introduced by approximation of the collision operator.

**1. Introduction**

A widely used tool for calculating neoclassical transport in stellarators is the DKES (Drift Kinetic Equation Solver) code [1,2]. Other codes such as PENTA [3,4] have been developed recently which use the calculations of DKES as input, compensating for the lack of momentum conservation by the model collision operator used in DKES [5]. In the drift-kinetic equation solved by DKES, not only is the collision operator simplified, but ad-hoc changes are also made to several other terms in order to facilitate rapid computation. These changes, which will be explained in more detail in section 2, are sometimes referred to as a "monoenergetic" approximation [6]. It is natural to ask what effect the monoenergetic approximation has on the transport coefficients. To rigorously assess the validity of this approximation in neoclassical calculations, it is necessary to solve the drift-kinetic equation with collisions or use a Monte-Carlo approach.

However, to avoid the complexity of the collision operator, a simpler calculation which is instructive is the determination of collisionless effective particle orbits. The characteristic curves of the kinetic equation represent effective particle trajectories, and so the effective trajectories associated with the DKES kinetic equation can be compared with the true particle trajectories. It has already been noted elsewhere that along the DKES trajectories, neither total energy nor magnetic moment are conserved [6]. Here, using a quasisymmetric vacuum magnetic field as a model, we will calculate the guiding-center trajectories analytically for both the equations used in DKES and for the true equations of motion. We will find that when there is a nonzero radial electric field, the trajectories used by DKES are fundamentally different than the true particle trajectories, and for both trajectories we will derive the shape of the trapped-passing boundaries in velocity space. It will be shown that DKES systematically under-predicts the trapped particle fraction.

The present analytical calculation complements and provides insight into a recent numerical study by Beidler et al [6]. In that work, the particle diffusion coefficient calculated by DKES was compared to Monte-Carlo calculations. It was found that the DKES diffusion coefficient can be substantially erroneous once the radial electric field $E_r$ approaches the "resonant" value $E_r^{res}$, at which the poloidal $\mathbf{E}\times\mathbf{B}$ drift cancels the poloidal component of parallel motion. We too will find that a fundamental problem arises in monoenergetic calculations when the magnitude of $E_r$ becomes of order $E_r^{res}$. These findings are relevant to the HSX stellarator [7], since it is expected that $E_r \sim E_r^{res}$ near the magnetic axis of this device [8], and so monoenergetic transport calculations are likely to be unreliable. In another recent work, Maassberg et al [5] examined the lack of momentum conservation by the collision operator in DKES and various compensation schemes. This investigation is complementary to the analysis herein, for the error in the trapped fraction discussed in the following sections is associated with the collisionless trajectories, and it is therefore unrelated to the fact that DKES uses an approximate collision operator rather than the full Fokker-Planck operator.

## 2. The monoenergetic approximation

Before specifying the monoenergetic equations used in DKES, we first review the "true" equations of guiding-center motion in steady electric and magnetic fields. First, $\dot{H} = 0$ where $H = v^2/2 + Ze\Phi/m$ is the total energy and the over-dot is a total time derivative $d/dt$, $m$ is the particle mass, $Ze$ is the particle charge, $\Phi$ is the electrostatic potential. Second, $\dot{\mu} = 0$ where $\mu = v_\perp^2/(2B)$ is the magnetic moment. The final dynamical equation is

$$\dot{\mathbf{r}} = v_\| \mathbf{b} + \mathbf{v}_m + \mathbf{v}_E, \tag{1}$$

where $\mathbf{b} = \mathbf{B}/B$, $\mathbf{v}_E = cB^{-2}\mathbf{B}\times\nabla\Phi$,

$$\mathbf{v}_m = \frac{v_\|}{\Omega}\nabla\times(v_\|\mathbf{b}) = \frac{v_\perp^2}{2\Omega B^2}\mathbf{B}\times\nabla B + \frac{v_\|^2}{\Omega B}\mathbf{B}\times(\mathbf{b}\cdot\nabla\mathbf{b}) + \frac{v_\|^2}{\Omega}\mathbf{b}\mathbf{b}\cdot\nabla\times\mathbf{b}, \tag{2}$$

$\Omega = ZeB/(mc)$ is the gyrofrequency, and the first curl in (2) holds $v$ and $\mu$ fixed. The parallel velocity correction in (2) (the last term) is consistent with [9], and it will be needed in section 4 to ensure conservation of helical momentum in the quasisymmetric magnetic field.

Now let us assume that the plasma consists of nested flux surfaced labeled by a radial coordinate $\psi$ satisfying $\mathbf{B}\cdot\nabla\psi = 0$, and let us take $\Phi$ to be a flux function. We assume MHD equilibrium so $(\nabla\times\mathbf{B})\cdot\nabla\psi = 0$. The radial equation of motion becomes $\dot{\psi} = (2v_\|^2 + v_\perp^2)(2\Omega B^2)^{-1}\mathbf{B}\times\nabla B\cdot\nabla\psi$.

The guiding-center equations of motion can be cast in terms of the speed $v$ and the pitch angle $p = v_\|/v$, with the result

$$\dot{v} = \frac{c\Phi' v}{2B^3}(1+p^2)\mathbf{B}\times\nabla\psi\cdot\nabla B \tag{3}$$

where the prime denotes $d/d\psi$, and

$$\dot{p} = -\frac{\upsilon}{2B^2}\left(1-p^2\right)\mathbf{B}\cdot\nabla B + \frac{c\Phi'}{2B^3}p\left(1-p^2\right)\mathbf{B}\times\nabla\psi\cdot\nabla B - \frac{\left(1-p^2\right)}{2pB}\mathbf{v}_{\mathrm{m}}\cdot\nabla B. \tag{4}$$

It is at this point that the ad-hoc changes are made to obtain the monoenergetic equations used in DKES. The magnetic drifts $\mathbf{v}_{\mathrm{m}}$ are neglected in the $\dot{\mathbf{r}}$ equation (1) and the $\dot{p}$ equation (4). Next, the $\Phi'$ term is dropped in (4), and (3) is replaced with $\dot{\upsilon}=0$. Lastly, the $\mathbf{E}\times\mathbf{B}$ drift in the $\dot{\mathbf{r}}$ equation is replaced with $\tilde{\mathbf{v}}_{\mathbf{E}} = c\Phi'\langle B^2\rangle^{-1}\mathbf{B}\times\nabla\psi$, which differs from the true $\mathbf{v}_{\mathbf{E}}$ in that $\nabla\cdot\tilde{\mathbf{v}}_{\mathbf{E}}=0$. The rationale for these approximations is to reduce the number of independent variables in the system by treating $\upsilon$ as a parameter, while maintaining a conservative structure of the equations.

To summarize, the effective equations of motion used in DKES are $\dot{\mathbf{r}} = \upsilon_\parallel \mathbf{b} + \tilde{\mathbf{v}}_{\mathbf{E}}$ and $\dot{p} = -(\upsilon/2)B^{-2}\left(1-p^2\right)\mathbf{B}\cdot\nabla B$ with $\upsilon$ a constant.

## 3. Monoenergetic trajectories

To examine how the particle trajectories are altered when these approximations are made, we consider a model magnetic field in which an analytic solution of the equations is possible, specified as follows. We consider poloidal and toroidal Boozer angles $\theta$ and $\zeta$ satisfying $\mathbf{B} = \nabla\psi\times\nabla\theta + \iota\nabla\zeta\times\nabla\psi$ with $2\pi\psi$ equaling the toroidal flux and $\iota(\psi)$ the rotational transform. We take $\mathbf{B}$ to be a vacuum field so $\mathbf{B} = I\nabla\zeta$, where $I$ is position-independent. We further assume that the field is quasisymmetric [10-12], and we ignore the radial variation in $B$ in the region of interest, so $B(\psi, \theta, \zeta) = B(\chi)$ where $\chi = M\theta - N\zeta$ is a helical angle. Finally, we take $\Phi'$ to be constant.

The DKES equations of motion can then be written as

$$\dot{p} = \frac{(N-\iota M)}{M}\frac{\upsilon}{2B}\left(1-p^2\right)\left(\frac{\partial B}{\partial \theta}\right)\mathbf{b}\cdot\nabla\zeta, \tag{5}$$

$$\dot{\zeta} = \upsilon_\parallel \mathbf{b}\cdot\nabla\zeta, \tag{6}$$

and

$$\dot{\theta} = \left[\iota\upsilon_\parallel + \frac{(N-\iota M)}{M}u\frac{B}{B_0}\right]\mathbf{b}\cdot\nabla\zeta, \tag{7}$$

where $B_0$ is the minimum of $B(\chi)$, and we have introduced the normalized radial electric field

$$u = \frac{M}{(N-\iota M)}\frac{B_0 cI\Phi'}{\langle B^2\rangle} \tag{8}$$

which is a constant of the motion. Several of the factors in this definition of $u$ have been chosen to simplify equations which follow.

From (6) and (7) we can also form

$$\dot{B} = \frac{(N-\iota M)}{M}\left(u\frac{B}{B_0} - \upsilon_\parallel\right)\left(\frac{\partial B}{\partial \theta}\right)\mathbf{b}\cdot\nabla\zeta. \tag{9}$$

Combining this result with (5), we obtain the following ordinary differential equation for $p(B)$:

$$(ub - vp)\frac{dp}{dB} = \frac{v}{2B}(1 - p^2) \qquad (10)$$

where $b = B/B_0$. This equation has the solution $(vp - ub)^2 - v^2 - u^2 b^2 = bX$ for some constant $X$ (as can be verified by differentiating the solution and then algebraically eliminating $X$.) Next we eliminate $X$ in favor of $v_{\|0}$, which is the value of $v_\|$ when the orbit crosses through the field minimum $B_0$. The result is

$$(v_\| - ub)^2 = (v_{\|0} - u)^2 b - v^2 b - u^2 b + v^2 + u^2 b^2. \qquad (11)$$

A particle is trapped if the right-hand side of this equation is negative when $b = b_{max}$. It follows that the trapped-passing boundary is described by

$$v_{\|0} = (B_{max}/B_0)u \pm v_{\perp 0}\sqrt{(B_{max}/B_0) - 1} \qquad (12)$$

where $v_{\perp 0}^2 = v^2 - v_{\|0}^2$.

If the variation in $B$ is weak, i.e. $\varepsilon \ll 1$ where $\varepsilon = B_{max}/(2B_0)$, we can write (12) as

$$v_{\|0} = u \pm v_{\perp 0}\sqrt{2\varepsilon}. \qquad (13)$$

This is precisely the usual $\Phi' = 0$ relation for particle mirroring, but with the apex of the passing "cones" shifted by $u$ in the parallel direction.

## 4. True trajectories

For the true trajectories, it is not $v$ but rather the total energy $H = v_\|^2/2 + \mu B + Ze\Phi/m$ which is constant. We eliminate $H$ in this relation in favor of quantities associated with a time when the trajectory passes through $B = B_0$, which will also denoted with a subscript 0:

$$\frac{v_\|^2}{2} + \mu B + \frac{Ze\Phi(\psi)}{m} = \frac{v_{\|0}^2}{2} + \mu B_0 + \frac{Ze\Phi(\psi_0)}{m}. \qquad (14)$$

To analyze this equation further, another relation between $\psi$ and $\psi_0$ is required. This relation can be obtained in a quasisymmetric field from the conservation of helical momentum. Using the original equations of motion (1)-(2) and the quasisymmetric field defined in section 3, it can be verified that $\dot{\psi}_* = 0$ where $\psi_* = M\psi_p - N\psi - MIv_\|/\Omega$ and $2\pi\psi_p$ is the poloidal flux. Proofs of the conservation of $\psi_*$ can be found in [10], [11], or appendix A of [13]. We can write $\psi_*$ in terms of values associated with a time at which the trajectory passes through $B = B_0$, which will also be denoted with a subscript 0:

$$(N - \iota M)\psi + \frac{MIv_\|}{\Omega} = (N - \iota M)\psi_0 + \frac{MIv_{\|0}}{\Omega_0}. \qquad (15)$$

As $\Phi(\psi) - \Phi(\psi_0) = (\psi - \psi_0)\Phi'$, we can use (15) to eliminate $\Phi(\psi)$ and $\Phi(\psi_0)$ in (14), giving

$$\frac{v_\|^2}{2} + \left(v_{\|0} - v_\|\frac{B_0}{B}\right)\eta u = \frac{v_{\|0}^2}{2} - (B - B_0)\mu \qquad (16)$$

where $\eta = \langle B^2 \rangle / B_0^2$ is a geometric coefficient which is close to 1. Solving (16) for $v_\|$ gives

$$v_\| = \frac{\eta}{b}u \pm \sqrt{\frac{\eta^2}{b^2}u^2 - 2\eta u v_{\|0} + v_{\|0}^2 - 2(B - B_0)\mu}. \qquad (17)$$

A particle is trapped if the radicand in (17) is negative for $B = B_{max}$, and so the trapped-passing boundary is obtained by setting the radicand to zero:

$$v_{\|0}^2 - 2\eta u v_{\|0} + \eta^2 \frac{B_0^2}{B_{max}^2} u^2 + \left(1 - \frac{B_{max}}{B_0}\right) v_{\perp 0}^2 = 0. \tag{18}$$

Rearranging,

$$v_{\|0} = \eta u \pm \sqrt{\left(\frac{B_{max}}{B_0} - 1\right) v_{\perp 0}^2 + \left(1 - \frac{B_0^2}{B_{max}^2}\right) u^2 \eta^2}. \tag{19}$$

In the $\varepsilon \ll 1$ weak-$B$-variation limit, this expression becomes

$$v_{\|0} = u \pm \sqrt{2\varepsilon}\sqrt{v_{\perp 0}^2 + 2u^2}. \tag{20}$$

Notice that these last two equations for the true trapped-passing boundaries differ from (12)-(13) for the corresponding boundary in DKES.

## 5. Trapped particle fraction

Figure 1 shows the trapped-passing boundaries for the two models. Both the DKES curve and the true curve show trapping near $v_{\|0} \approx u$, but the true trapped region is larger than the DKES prediction. If the radial electric field vanishes ($u = 0$), then the DKES equations reduce to the correct equations of motion (to leading order in the ratio of gyroradius to system size), and so DKES gives the correct prediction for the trapped-passing boundary.

To describe the difference in particle trapping between DKES and the true equations of motion, it is useful to define $f_t$, the fraction of a stationary Maxwellian distribution at $B = B_0$ which would be trapped:

$$f_t = \frac{1}{v_{th}^3 \sqrt{\pi}} \int_0^\infty dv_{\perp 0}^2 \int dv_{\|0} \exp\left(-\frac{v_{\perp 0}^2 + v_{\|0}^2}{v_{th}^2}\right) \tag{21}$$

where $v_{th} = \sqrt{2T/m}$ is the thermal speed, and the range of the $v_{\|0}$ integration is taken from (12), (13), (19), or (20). For $\varepsilon \ll 1$, we can approximate the last $v_{\|0}^2$ in (21) by $u^2$, allowing the integrals to be performed analytically, giving

$$f_t = \begin{cases} \sqrt{2\varepsilon} e^{-U^2} & \text{(monoenergetic)} \\ \sqrt{2\varepsilon} e^{-U^2}\left(2\sqrt{\frac{2}{\pi}}|U| + \left[1 - \text{erf}\left(\sqrt{2}|U|\right)\right] e^{2U^2}\right) & \text{(actual)} \end{cases} \tag{22}$$

where $U = u/v_{th}$. To leading order in $\varepsilon$, $U = E_r / E_r^{res}$ if $U$ is defined with the ion mass and where the "resonant" electric field $E_r^{res}$ is defined in [6,8]. The functions in (22) are plotted in Figure 2a. Note that $U \propto \sqrt{m}$, so the effects calculated herein may be important for ions but not for electrons.

## 6. Discussion

Both the monoenergetic and true equations imply that there is an exponential ($\exp(-U^2)$) decrease in the trapped fraction as $E_r$ increases. Figure 2 shows that when $u \neq 0$, DKES systematically underestimates the trapped particle fraction. Figure 2b shows the ratio of the two functions in (22), showing that the discrepancy is as large as a factor of 2 for $U>1$. As noted in the introduction, it has been predicted that $U>1$ for ions in the core of HSX [8], which is nearly quasisymmetric, and so the analysis herein has bearing on that experiment.

As trapped particles play a central role in the physics of neoclassical phenomena, it is reasonable to expect that the effect shown in figure 2b will distort the results of monoenergetic neoclassical calculations when $U \sim 1$. The bootstrap current, radial fluxes, and Hirshman-Sigmar viscosity coefficients in a quasisymmetric, high-aspect-ratio, low-collisionality plasma with $E_r \ll E_r^{\text{res}}$ are proportional to an *effective* trapped fraction

$$f_t^{\text{eff}} = 1 - \frac{3}{4}\langle B^2 \rangle \int_0^{1/B_{\max}} \frac{\lambda\, d\lambda}{\langle \sqrt{1-\lambda B} \rangle} \qquad (23)$$

which differs from the $f_t$ defined in (21). However, it is not valid to simply replace $f_t^{\text{eff}}$ in the expressions for the viscosity coefficients or other neoclassical quantities by the second line of (22). To evaluate neoclassical fluxes and flows properly, it is necessary to solve the drift-kinetic equation with collisions. This problem is analyzed for a quasisymmetric field in [13], using a momentum-conserving model collision operator to obtain corrected expressions.

A magnetic field cannot be perfectly quasisymmetric, so accurate evaluation of neoclassical transport in an experimentally relevant magnetic field requires numerical computation. However, as the calculation herein shows, when $E_r$ approaches $E_r^{\text{res}}$, the radial magnetic drift causes significant changes to a particle's energy through variation in the electrostatic potential, even when this potential is a flux function. Thus, proper evaluation of neoclassical transport in the $E_r \sim E_r^{\text{res}}$ regime likely requires a code which is radially nonlocal so this radial magnetic drift can be retained.

## Acknowledgements

This research was supported by the United States Department of Energy under grant DE-FG02-91ER-54109.

**Figure 1**

(Colour online) The trapped region of velocity space at a location where $B = B_0$. (a) The monoenergetic equations give the correct trapping boundary when there is no electric field ($u = 0$). (b) When $u \neq 0$, the actual trapped region is larger than the one obtained from the monoenergetic equations.

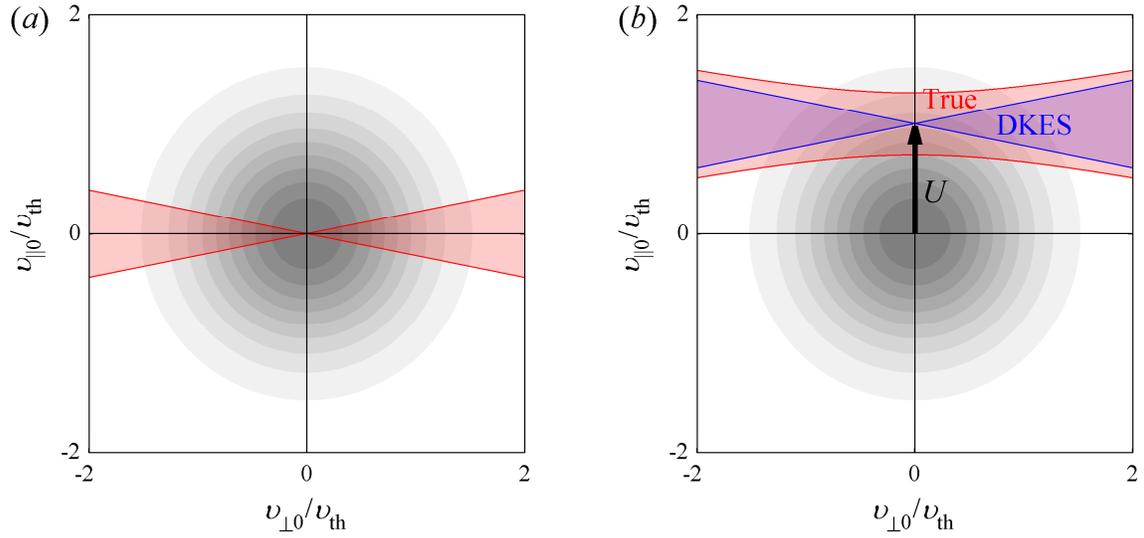

**Figure 2**

(Colour online) (a) Trapped particle fraction in the two models, eq. (25). (b) Ratio of the monoenergetic trapped fraction divided by the actual trapped fraction.

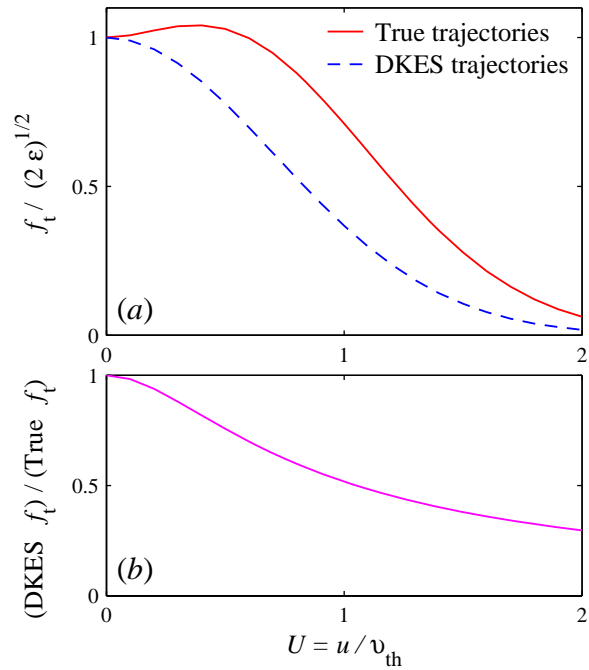